\documentclass[12pt]{article}

\makeatletter\renewcommand{\section}{\@startsection
	{section}{1}{\z@}{-3.5ex plus -1ex minus
		-.2ex}{2.3ex plus .2ex}{\bf }}

\makeatletter\renewcommand{\subsection}{\@startsection{subsection}{2}{\z@}{-3.25ex
		plus -1ex minus
		-.2ex}{1.5ex plus .2ex}{\it }}
\makeatletter\renewcommand{\subsubsection}{\@startsection{subsubsection}{3}{-2.45ex}{-3.25ex
		plus -1ex minus -.2ex}{1.5ex plus .2ex}{\it }}

\makeatletter \@addtoreset{equation}{section}

\usepackage{graphicx}
\graphicspath{{plots/}} 
\usepackage{float} 
\usepackage{epsfig} 
\usepackage{verbatim}
\usepackage{hyperref}
\usepackage{easyfig}
\usepackage{bbm}
\usepackage{amsmath}
\usepackage{amsfonts}
\usepackage[compact]{titlesec}
\usepackage{subdepth}
\usepackage{dsfont}
\usepackage{braket}
\usepackage{tikz}
\usepackage[utf8]{inputenc}
\usepackage[T1]{fontenc}
\usepackage{amsmath}
\usepackage[toc]{glossaries}

\usepackage{caption}
\usepackage{subcaption}

\renewenvironment{thebibliography}[1]
{\baselineskip=16pt plus 2pt minus 1pt
	\section*{\large\refname
		\@mkboth{\MakeUppercase\refname}{\MakeUppercase\refname}}%
	\list{\@biblabel{\@arabic\c@enumiv}}%
	{\settowidth\labelwidth{\@biblabel{#1}}%
		\leftmargin\labelwidth
		\advance\leftmargin\labelsep
		\@openbib@code
		\usecounter{enumiv}%
		\let\p@enumiv\@empty
		\renewcommand\theenumiv{\@arabic\c@enumiv}}%
	\sloppy
	\clubpenalty4000
	\@clubpenalty \clubpenalty
	\widowpenalty4000%
	\sfcode`\.\@m}

\let\fn\footnote
\renewcommand{\footnote}[1]{\linespread{1.1}\fn{#1}\linespread{1.29}}

\usepackage{physics}
\usepackage{amsfonts}
\usepackage{amssymb}
\usepackage{mathrsfs}
\usepackage{amsmath,amssymb}
\usepackage{bm}
\usepackage{caption}
\usepackage{amsmath}
\usepackage{graphicx}
\usepackage{cite}

\usepackage[T1]{fontenc}
\usepackage[utf8]{inputenc}

\def\tyng(#1){\hbox{\tiny$\yng(#1)$}}

\newcommand{\be}{\begin{equation}}
	\newcommand{\ee}{\end{equation}}
\newcommand{\bea}{\begin{array}}
	\newcommand{\ea}{\end{array}}
\newcommand{\beqa}{\begin{eqnarray}}
	\newcommand{\eeqa}{\end{eqnarray}}

\usepackage{pdfpages}

\begin{document}



	\begin{titlepage}
	\begin{flushright}
	\end{flushright}
	
	
	\begin{center}
		{\Large \bf Review of Twisted Poincar\'e Symmetry }\\
		~\\
		
		
		\vskip 3em
		
		\centerline{$ \text{\large{\bf{A.P.Balachandran}}} \, \, $, $ \text{\large{\bf{S. K\"{u}rk\c{c}\"{u}o\v{g}lu}}}$, $\text{\large{\bf{S. Vaidya}}} $ }
		\vskip 0.5cm
		\centerline{\sl Department of Physics, Syracuse University, }
		\centerline{\sl Syracuse NY, 13244, USA }
		\centerline{\sl Department of Physics, Middle East Technical University,}
		\centerline{\sl Dumlupınar Boulevard, 06800, Ankara, Turkey}
		\centerline{\sl Centre for High Energy Physics, Indian Institute of Science,}
		\centerline{\sl Bengaluru 560012, India }
		
		\vskip 1em
		
		\begin{tabular}{r l}
			E-mails: 
			&\!\!\!{\fontfamily{cmtt}\fontsize{11pt}{15pt}\selectfont aibalach@syr.edu}\\
			&\!\!\!{\fontfamily{cmtt}\fontsize{11pt}{15pt}\selectfont kseckin@metu.edu.tr}  \\
			&\!\!\!{\fontfamily{cmtt}\fontsize{11pt}{15pt}\selectfont vaidya@iisc.ac.in}
			
		\end{tabular}
		
	\end{center}
	
	\vskip 5 em
	
	\begin{quote}
		\begin{center}
			{\bf Abstract}
		\end{center}
		
		This article reviews the construction and some applications of twisted Poincare-covariant quantum fields on the Moyal plane. The Drinfel'd twist, which plays a key mathematical role in this construction, is then applied to the case of discrete groups, with a view to applications to geons in quantum gravity. The Poincar\'e-twisted fields can also be applied to study the CMB anisotropies, and corrections to the power spectrum are used to put constraints on spacetime noncommutativity. The article also addresses the issue of the difference between Moyal and Voros quantum fields. Finally, it is pointed out that the Euclidean functional integrals of QFTs on the Moyal plane do not, in general, obey reflection positivity.

		\vskip 1em

	\end{quote}
	
\end{titlepage}

\setcounter{footnote}{0}
\pagestyle{plain} \setcounter{page}{2}

\newpage

\section{Introduction}

One of the most compelling arguments for the emergence of noncommutative spacetimes at length scales close to the 
Planck length $l_P$ comes from the work of Doplicher, Fredenhagen and Roberts (DFR) \cite{DFR, Doplicher}. Intuitively, their argument 
can be summarized as follows. Conventionally, events are taken to be points of a commutative manifold, the Minkowski 
spacetime $M^{3,1}$. However, attempts to localize (i.e. measure) events with extreme precision require the use of probes 
with arbitrarily high energies concentrated in arbitrarily small spacetime regions. But according to classical 
general relativity, such large concentrations of energy can lead to the gravitational collapse and subsequent formation of 
event horizons. Hence localization of events with arbitrary precision has no operational meaning. 

DFR then argue that the commutative spacetime $M^{3,1}$ should be replaced by a spacetime possessing a quantum 
structure, or a noncommutative spacetime, wherein the spacetime uncertainty relations are straightforward consequences.
Specifically, they argue that spacetime at short length scales should be described by a noncommutative algebra ${\cal E}$, 
and the points by pure states on ${\cal E}$. Also, the commutative manifold $M^{3,1}$ should emerge from ${\cal E}$ at large distances, i.e. distances much larger than $l_P$. We will briefly review the arguments of DFR \cite{DFR, Piacitelli} below.

Consider a free neutral scalar field $\phi$ in a state $\Phi$ given by $\Phi = e^{i \phi(f)} \Omega$, where $\Omega$ is the vacuum state vector, and $f$ a real smooth test function with support in a compact region whose extent is given by $\Delta x_\mu$. The Hamiltonian, derived from the $T_{00}$ component of the stress-energy tensor $T_{\mu \nu}$ can be used to make a heuristic estimate of the energy $E$ concentrated in the region $(\Delta x_1,\Delta x_2,\Delta x_3)$, with temporal uncertainty $\Delta x_0 \sim 1/E$. The gravitational potential at $x_\mu \simeq 0$, computed from the retarded potential is then required to obey the condition that photons of energy $\epsilon$ should {\it not} be trapped. This leads to the uncertainty relations
\begin{eqnarray}
\Delta x_0 \sum_{i=1}^3 \Delta x_i &\gtrsim& l_P^2 \,,
\label{unct1} \\
\Delta x_j \Delta x_k &\gtrsim& l_P^2 \,.
\label{stuncertain}
\end{eqnarray}
Finally, DFR argue that the noncommutative algebra 
\begin{equation}
[q_\mu , q_\nu] = \theta_{\mu \nu}, \quad [q_\lambda, \theta_{\mu \nu}]=0
\label{ncst}
\end{equation}
yields the uncertainty relations in (\ref{unct1}) and (\ref{stuncertain}).
 
Corresponding to the algebra (\ref{ncst}), Bahns, Doplicher, Fredenhagen and Piacitelli  \cite{Piacitelli} define operators for 
areas, 3-volumes and 4-volumes \cite{Piacitelli}, which we will describe next very briefly. Starting with the universal 
enveloping algebra of (\ref{ncst}),  authors of \cite{Piacitelli} define a universal differential calculus, allowing the construction of the analogs of 1-, 2-, and 3-forms. The operators corresponding to area, 3-volume and 4-volume have interesting and unexpected properties. For the area operator, the sum of the modulus squared of the components is bounded below. The spectrum of the 3-volume operator is $\mathbb{C}$, the complex plane. The 4-volume operator has a pure point spectrum, and again the sum of the modulus squared of its components is bounded below.

The physical consequences of these remarkable properties described above have not been sufficiently explored.

We next describe the construction of covariant fields on the Moyal plane.

\section{Covariant Quantum Fields on the Moyal Plane}

The algebra of smooth functions with values in $\mathbb{C}$, ${\cal A}_0(M)$ on a manifold $M$ is a commutative algebra under pointwise multiplication. It is possible to recover the topology, and even the differential structure of $M$ using the results of Gel'fand and Naimark, and subsequent far-reaching results of Connes and coworkers \cite{Connes1, Connes2}. It is this algebra that describes the "configuration space" in quantum mechanics.

The Moyal plane ${\cal A}_\theta(M)$ is a noncommutative deformation of ${\cal A}_0(M)$. If $m_0$ is the multiplication map of ${\cal A}_0(M)$, $m_0(f_1 \otimes f_2)(x) = f_1(x) f_2(x)$ for $f_i \in {\cal A}_0(M), x \in M$, then the twisted multiplication map $m_\theta$ for ${\cal A}_\theta(M)$ is
\begin{equation}
m_\theta(f_1 \otimes f_2)(x) = (f_1 e^{\frac{i}{2} \overleftarrow{\partial}_\mu \theta_{\mu \nu} \overrightarrow{\partial}_\nu} f_2)(x)
\end{equation}
where $\theta_{\mu \nu}$ is a constant antisymmetric matrix. This can also be written as 
\begin{equation}
m_0(F_\theta (f_1 \otimes f_2)), \quad F_\theta = e^{\frac{i}{2} \theta_{\mu \nu} \partial_\mu \otimes \partial_\nu}.
\end{equation}
The operator $F_\theta$ is called the Drinfel'd twist \cite{drinfeld, chaichian,wess, Aschieri1, Aschieri2, sasha}.

We will write the product of functions in ${\cal A}_\theta(M)$ with the notation $*_\theta$. Thus $(f_1 *_\theta f_2)(x) = (f_1 e^{\frac{i}{2} \overleftarrow{\partial}_\mu \theta_{\mu \nu} \overrightarrow{\partial}_\nu} f_2)(x)$.

The product $*_\theta$ is associative but not commutative. These are best seen by using plane waves $e_p$, $e_p(x) = e^{i p x}$ for which $P_\mu e_p = p_\mu e_p$, $P_\mu$ being the translations. Then $e_p *_\theta e_q = e^{-\frac{i}{2} p \wedge q} e_p e_q, p \wedge q := p_\mu \theta_{\mu \nu} q_\nu$. Associativity follows easily from this formula. But since $e_q *_\theta e_p = e^{ i/2 p \wedge q} e_q e_p \neq e_p *_\theta e_q$, it is not commutative. 

The action of the Poincar\'e group ${\cal P}_+^\uparrow = \{(a, \Lambda) \}$ on ${\cal A}_0(M)$ is $((a, \Lambda)f)(x) :=[(a,\Lambda)\triangleright f](x)= f(a, \Lambda)^{-1} x)$. Its action on the product of functions in ${\cal A}_0(M)$ is well-known. We can write it using the canonical coproduct $\Delta_0$ on ${\cal P}_+^\uparrow: \Delta_0 ((a, \Lambda)) = (a,\Lambda) \otimes (a,\Lambda)$. Then 
\begin{equation}
(a,\Lambda) \vartriangleright (f_1 \otimes f_2)=m_0(\Delta_0(a,\Lambda) \vartriangleright (f_1 \otimes f_2)) := 
m_0([(a,\Lambda) \vartriangleright f_1] \otimes [(a,\Lambda) \vartriangleright f_2])
\end{equation}
But this action is not compatible with $m_\theta$: $m_\theta[\Delta_0(a,\Lambda) \vartriangleright (f_1 \otimes f_2)] \neq (a,\Lambda) (f_1 *_\theta f_2)$. However, the Drinfel'd-twisted coproduct $\Delta_\theta$, $\Delta_\theta[(a,\Lambda)]= F_\theta^{-1}[(a,\Lambda) \otimes (a,\Lambda)]F_\theta$, is compatible with $m_\theta: m_\theta[\Delta_\theta(a,\Lambda) \vartriangleright (f_1 \otimes f_2)]= (a,\Lambda)\vartriangleright (f_1 *_\theta f_2)$. This coproduct is co-commutative \cite{chari,majid,aschieri,chaichian,wess, Aschieri1,Aschieri2}. 

It is this action of ${\cal P}_+^\uparrow$ that we want to adapt to quantum fields and generalize also to commutative discrete subgroups of a group $G$ acting on quantum fields. 

There is an elegant way to twist the functions belonging to ${\cal A}_0(M)$ to operator-valued maps $f_\theta$ which incorporate the twisted product. The definition of $f_\theta$ is
\begin{equation}
f_\theta = f e^{\frac{i}{2}\overleftarrow{\partial_\mu} \theta_{\mu \nu} \overrightarrow{\partial_\nu}}
\end{equation}
Such a twist was introduced by Grosse \cite{Grosse} and Zamalodchikovs and Faddeev \cite{Zamo,faddeev} in the context of integrable models. We can easily check that 
\begin{equation}
f_{1 \theta} f_{2 \theta} = (f_1 *_\theta f_2)_\theta, \quad f_{i \theta} = 
f_i e^{\frac{i}{2}\overleftarrow{\partial}_\mu \theta_{\mu \nu} \overrightarrow{\partial}_\nu}
\end{equation}
For plane waves, we have
\begin{equation}
e_{p,\theta} = e_p e^{-\frac{1}{2} p \wedge \overrightarrow{\partial}}
\end{equation}
The algebra of $f_\theta$'s is isomorphic to ${\cal A}_\theta(M)$ and we will use the same name for both.

We note that $e_{p,\theta} e_{q,\theta} = e_{p+q, \theta}$ where there is still the exponential with $\partial_\mu$ in the right extreme. We can get rid of it by applying $e_{p+q, \theta}$ (and polynomials of $e_{p,\theta}$) on the constant function ${\mathbf 1}$ with value 1. Then the Lorentz group ${\cal P}_+^{\uparrow}$ with the coproduct $\Delta_\theta$ acts consistently on $e_p$'s and hence $f_\theta$'s: $\Delta_\theta (a, \Lambda) \triangleright e_{p,\theta} \cdot  e_{q,\theta} {\mathbf 1}=e_{\Lambda p,\theta} e_{\Lambda q, \theta} e^{i/2(\Lambda p \wedge \Lambda q) \cdot a} {\mathbf 1}$. It is this 
``dressed'' approach that we will generalize to quantum fields. The role of ${\mathbf 1}$ is taken in that case by the vacuum state.

Next suppose that $\phi$ is a covariant scalar quantum field. The Poincar\'e group ${\cal P}_+^\uparrow$ acts on $\phi$ as it did on ${\cal A}_0(M)$ and is represented by the unitary operator $U(a, \Lambda)$ if $(a,\Lambda) \in {\cal P}_+^\uparrow$. Thus we have $U(a,\Lambda) \phi(x) U(a,\Lambda)^{-1} = \phi(\Lambda x + a)$. Its action on products of fields is also given by such conjugations. 

Let us consider free, in or out fields of mass $m$ so that we can write 
\begin{eqnarray}
\phi_0(x) &=& \int \frac{d^3p}{2p_0} (a_p e_p(x) + a_p^* e_{-p}(x)), \quad p_0 = \sqrt{\vec{p}\,^2 + m^2}, \nonumber \\ 
&:=& \phi^+(x) + \phi^ -(x). 
\end{eqnarray}
On the vacuum,
\begin{equation}
\phi(x) |0\rangle = \phi_-(x) |0\rangle
\end{equation}
and $U(a,\Lambda)\phi_-(x)|0\rangle = \phi_-(\Lambda x +a)|0\rangle$ since $U(a,\Lambda)|0\rangle = |0\rangle$.
We have similar formulae for products of several fields. 

The field $\phi_\theta$ twisted by $F_\theta$ is
\begin{equation}
\phi_\theta(x)= \phi_0 e^{-i/2 \overleftarrow{P_\mu} \theta_{\mu \nu} \overrightarrow{P_\nu}}
\end{equation}
so that
\begin{eqnarray}
\phi_\theta (x) |0 \rangle &=& \phi_0 (x) |0 \rangle, \\
\phi_\theta (x_1) \phi_\theta (x_2) |0 \rangle &=& \phi_0 (x) e^{-\frac{1}{2}\overleftarrow{P}_\mu \theta_{\mu \nu} \overrightarrow{P}_\nu} \phi_0 (x_2) |0 \rangle
\end{eqnarray}
etc.

We can write 
\begin{equation}
\phi_\theta(x_1)\phi_\theta(x_2)|0\rangle = \int \prod_i d\mu(p_i) F_\theta(a^*_{p_i} \otimes a^*_{p_2}) |0\rangle e_{-p_1}(x_1) e_{-p_2}(x_2)
\end{equation}
where now $F_\theta = e^{-i/2 P_\mu \otimes \theta_{\mu \nu} P_\nu}$.
Consider the twisted coproduct 
\begin{equation}
\Delta_\theta(a,\Lambda) = F_\theta \lbrace (a,\Lambda) \otimes (a,\Lambda) \rbrace F^{-1}_\theta
\end{equation}
where we have not used a new notation for $F_\theta$ here. Then
\begin{equation}
\Delta_\theta[U(a,\Lambda)] = F_\theta U(a,\Lambda) \otimes U(a,\Lambda) F^{-1}_\theta \,.
\end{equation}
With this coproduct, ${\cal P}_+^\uparrow$ acts covariantly on the left-hand side:
\begin{equation}
\Delta_\theta[U(a,\Lambda)] \phi_\theta(x_1)\phi_\theta(x_2)|0\rangle = \phi_\theta(\Lambda x_1 +a) \phi_\theta(\Lambda x_2 +a ) |0\rangle
\end{equation}
as a short calculation shows. 

The coproduct $\Delta_0$ is compatible with symmetrization. Thus if $\sigma_0$ is the symmetrization operator 
for the $\phi_0$'s,
\begin{equation}
\sigma_0 \phi_0(x_1) \phi_0(x_2) |0 \rangle = \phi_0(x_2) \phi_0(x_1) |0\rangle
\end{equation}
then $\Delta_0[U(a,\Lambda)]\sigma_0 = \sigma_0\Delta_0[U(a,\Lambda)]$. This formula generalizes to products of fields as well, as discussed in \cite{balgroup}. For twisted fields, we have to twist $\sigma_0$ to 
\begin{equation}
\sigma_\theta = F_\theta \sigma_0 F^{-1}_\theta = F^2_\theta \sigma_0
\end{equation}
Note that $\sigma_\theta^2 = {\mathbf 1}$. This identity along with other relations defining the braid group are valid for action on polynomials of fields so that the braid group ${\cal B}_N$ becomes the permutation group $S_N$ for $N$ fields. We can then work with representations of $S_N$ on $\phi_\theta (x_1)\cdots \phi_\theta(x_N) |0\rangle$ without spoiling Poincar\'e covariance. We can work with Bose and para fields. 

Thus we see that twisted fields are also compatible with the (twisted) action of $S_N$.

We call quantum fields with a consistent action of ${\cal P}_+^\uparrow$ and $S_N$ as Poincar\'e covariant fields. Thus $\phi_0$ and $\phi_\theta$ are Poincar\'e covariant fields. Note that $\phi_\theta$ depends on an abelian subgroup of translations of the Poincar\'e group ${\cal P}_+^\uparrow$.

We want to generalize these considerations with ${\cal P}_+^\uparrow$ replaced by a generic group $G$ acting on quantum fields, and translations replaced by a {\it discrete} subgroup $H$ of $G$. We lead up to it by writing $F_\theta$ for ${\cal P}_+^\uparrow$ using projection operators. 

The operators $a_p,a^*_q$ have the commutation rules $[a_p, a_q^\dagger] = 2p_0 \delta^3(p-q)$. Hence if $|p \rangle = a_p^\dagger |0 \rangle$,
\begin{equation}
|p \rangle \langle p| \int d \mu(q) f(q) a_q^\dagger |0 \rangle = f(p) |p \rangle, \quad d\mu(q) = \frac{d^3 q}{2 q_0} \,.
\end{equation}

Let us write down the projection operator  as ${\cal P}_p$. Then ${\cal P}_p \otimes_S {\cal P}_q$, the symmetrised tensor product, projects a two-particle state $\int d\mu(k) d\mu(l) f(k,l) a^\dagger_k a^\dagger_l |0 \rangle $ to the two-particle subspace with momenta $p,q$:
\begin{equation}
({\cal P}_p \otimes_S {\cal P}_q) \int d\mu(k) d\mu(l) f(k,l) a^\dagger_k a^\dagger_l |0 \rangle = f(p,q) a^\dagger_p a^\dagger_q |0 \rangle \,.
\end{equation}
Such projectors extend to ${\mathbf 1}^{(N)}$, the projectors to the $N$-particle subspace.  We have,
\begin{equation}
\int d \mu(p) |p\rangle \langle p | = {\mathbf 1}^{(1)}, \quad \int d \mu(p) d \mu(q) |p,q \rangle \langle p,q | = {\mathbf 1}^{(2)} \,, \quad \mbox{etc} \,.
\end{equation}

We see also that 
\begin{equation}
\phi^-_\theta(x_1) \phi^-_\theta(x_2) |0 \rangle = \int d \mu(p) d \mu(q) e^{i p \wedge q} |p,q \rangle \langle p,q| \phi^-_\theta (x_1) \phi^-_\theta (x_2) |0 \rangle \,,
\end{equation}
which for $\theta = 0$ is just $\phi^-_0(x_1) \phi^-_0(x_2) |0\rangle$.

The action of the Poincar\'e group can also be written down:
\begin{multline}
\Delta_\theta(U_\theta(a,\Lambda)) \, (\phi^-_\theta(x_1) \phi^-_\theta(x_2) |0 \rangle )= F_\theta (U_0(a,\Lambda) \otimes U_0(a,\Lambda))F^{-1}_\theta \, (\phi^-_\theta(x_1) \phi^-_\theta(x_2) |0 \rangle ) \\
= \int d \mu(p) d \mu(q)  e^{i (\Lambda p) \wedge (\Lambda q)} e^{i (\Lambda p) a + i (\Lambda q) a} |\Lambda p, \Lambda q \rangle \langle p,q| \phi_\theta(x_1) \phi_\theta(x_2) |0\rangle \,.
\end{multline}

Important ingredients in these constructions are the projection operators.

Twisted quantum fields have implications for the spin-statistics relation, especially at high energies. They are also able to avoid UV-IR mixing, and hence more appropriate for making realistic models of particle physics. The details may be found in our works 
\cite{mangano, vaidya,BaPiQur,BaPiQue,deQueiroz:2012ti,Basu:2010qm,TRG,BAP,BP,Balachandran:2006ib}.

\subsection{The Treatment of Discrete Groups}

Actions of discrete groups $G$ on quantum fields occur extensively in the treatment of mapping class groups of the diffeomorphism groups in quantum gravity.

Mapping class groups are also called large gauge transformations. For ${\mathbb R}^3$, they are trivial, but they are not for more complicated asymptotically flat manifolds. An example is the 3-torus with a point (representing spatial infinity) removed. For reviews, we refer to \cite{Aneziris:1989cr}.

The mapping class groups $G$ are discrete, but generally non-abelian. We assume that we have a quantum field $\psi_0$ on the spatial slice with an action $G \ni g \triangleright \psi_0$ of $G$ on $\psi_0$. Here $g \triangleright \psi_0(p)= \psi_0(g^{-1}p)$ with $p$ being a point on this slice and $p \rightarrow g^{-1}p$ the action of $G$ on the manifold. This is the analog of the action of ${\cal P}_+^\uparrow$ on $\phi_0$. We want to twist the product of $\psi_0$'s and this action. The twist is on an abelian subgroup $A \subset G$, just as it was on the abelian translations contained in the Poincar\'e group for the case of ${\mathbb R}^4$.

To begin with, we assume that $A$ is finite. Then it is known that $A$ is the product of cyclic groups: $A= {\mathbb Z}_{n_1} \times {\mathbb Z}_{n_2} \times \cdots {\mathbb Z}_{n_k}$, where $n_j$ is the order of the cyclic group ${\mathbb Z}_{n_j}$.

The group ${\mathbb Z}_{n_j}$ has $n_j$ irreducible representations $\rho_{m_j}, m_j = 0,1,\cdots n_j-1$, where 
$\rho_{m_j}: z \in {\mathbb Z}_{m_j} \rightarrow \rho_{m_j}(z) = z^{m_j}$. The corresponding character $\chi_{m_j}$ is $\chi_{m_j}(z^k) = z^{k m_j}$.

The projection operator, the analogue of ${\cal P}_p$, acting on $\psi_0$ to project to the space for the action of $\rho_{m_j}$ is, in a bra-ket notation, then 
\begin{equation}
{\cal P}_{m_j} = \sum_{k_j=0}^{m_j -1} |m_j, k_j \rangle \langle m_j, k_j| 
\end{equation}
since 
\begin{eqnarray}
\hat{z} {\cal P}_{m_j} &=& \chi_{m_j}(z) {\cal P}_{m_j}, \\
{\cal P}_{m_j} {\cal P}_{m_j} &=& \delta_{m_j,m_k} {\cal P}_{m_j}.
\end{eqnarray}
as can easily be checked. 

Consider then the projection operator for the representation $ \rho_{\vec{m}} := {\cal \rho}_{m_1} \otimes {\cal \rho}_{m_2} \otimes \cdots \otimes {\cal \rho}_{m_k}$ of $A^{\otimes k }$. This projection operator is given as
\begin{equation}
{\cal P}_{\vec{m}} = {\cal P}_{m_1} \otimes {\cal P}_{m_2} \cdots \otimes {\cal P}_{m_k}, \quad \vec{m} = (m_1,m_2,\cdots m_k).
\end{equation}
In particular, the projection operator ${\cal P}_{\vec{m}} \otimes {\cal P}_{\vec{m}}$, projects to the representation  $\rho_{\vec{m}} \otimes \rho_{\vec{m}}$ of $A \otimes A$.

The untwisted multiplication map for the fields $\psi_0$ is 
\begin{equation}
m_0 (\psi_0 \otimes \psi_0)(p) = \psi_0 (p) \, \psi_0(p).
\end{equation}
We can now twist $m_0$ to $m_\theta$ using the abelian algebra $A$:
\begin{eqnarray}
m_\theta (\psi_0 \otimes \psi_0) &=& m_0[F_\theta (\psi_0 \otimes \psi_0)], \\
F_\theta &=& \sum_{\vec{m},\vec{m}'} e^{\frac{i}{2} m_i \theta_{ij} m'_j} {\cal P}_{\vec{m}} \otimes {\cal P}_{\vec{m}'}
\end{eqnarray}
In the above expression for $F_\theta$,  $\theta_{ij}$ represents an antisymmetric matrix with constant entries.

The action of $G$ on $\psi$ has also to be twisted as in the Poincar\'e case. If $G$ acts on $\psi_0 \otimes \psi_0$ using the canonical coproduct $\Delta_0$,
\begin{eqnarray}
\Delta_0(\hat{g}) = \hat{g} \otimes \hat{g} \\
\Delta_0(\hat{g})\triangleright (\psi_0 \otimes \psi_0) (p) &=& \psi_0(g^{-1}p) \psi_0(g^{-1}p) \,,
\end{eqnarray}
it now acts with a twisted coproduct $\Delta_\theta$:
\begin{equation}
\Delta_\theta(\hat{g}) = F_\theta \Delta_0 (\hat{g} \otimes \hat{g}) F^{-1}_\theta.
\end{equation}
Using ${\cal P}_{\vec{m}} {\cal P}_{\vec{m}'} = \delta_{\vec{m},\vec{m}'} {\cal P}_{\vec{m}}$, we can check that 
\begin{equation}
F^{-1}_\theta= \sum_{\vec{m},\vec{m}'}e^{-\frac{i}{2} m_i \theta_{ij} m_j} \otimes {\cal P}_{\vec{m}'} \,,
\end{equation}
and
\begin{equation}
\Delta_\theta(\hat{g}) \triangleright F_\theta((\psi_0 \otimes \psi_0) = F_\theta (\hat{g} \otimes \hat{g})(\psi_0 \otimes \psi_0) \,,
\end{equation}
the compatibility condition of the twisted coproduct with the twisted multiplication.

We remark that if $e^{\frac{i}{2} m_i \theta_{ij} m'_j} = 1$ for all $\vec{m},\vec{m}'$, $F_\theta= F_0$ and we can recover the untwisted case.

The next item is the twisted or the dressed field $\psi_\theta$. It is
\begin{equation}
\psi_\theta = \psi_0 \sum_{\vec{m},\vec{m}'} \overleftarrow{\cal P}_{\vec{m}} e^{-\frac{i}{2} m_i \theta_{ij} m'_j} \otimes \overrightarrow{\cal P}_{\vec{m}'}
\end{equation}
The vacuum of the Poincar\'e group is now replaced by a $G$-invariant vector which is assumed to exist. The rest of the discussion is as in the case of ${\cal P}_+^\uparrow$.

There are cases where $G$ contains ${\mathbb Z}$. So $A$ can contain one or more factors of ${\mathbb Z}$. We briefly examine what happens to our considerations if $A = {\mathbb Z}_n \times {\mathbb Z}$. Generalizations to several factors of cyclic groups or ${\mathbb Z}$'s are straightforward.

The irreducible representations of ${\mathbb Z}_n$ are $\rho_j, j=0,1,\cdots n-1$, $\rho_j(z) = z^j$, while those of ${\mathbb Z}$ are $\rho_\sigma$, $\rho_\sigma(m) = e^{i \sigma m \phi}$. Here $\phi$ are the coordinates on $S^1$, and $\sigma$ and $m$ are integers.

The projection operator to $\rho_j$ is ${\cal P}_j, j=0,1,\cdots n-1$ and explicitly given as ${\cal P}_j = \sum_{k=0}^{n-1} \bar{\chi}_j (z^k) \hat{z}^k$. The projection operator to the representation $\rho_\sigma$ is  ${\cal P}_\sigma$  where
\begin{equation}
({\cal P}_\sigma) f (\phi) = \frac{1}{2\pi} \int_0^{2\pi} d\phi' e^{i \sigma (\phi - \phi')} f(\phi') 
\end{equation}
where $f$ is a function on $S^1$. We can check that if $f(\phi) = \sum f_n e^{i n \phi}$,
\begin{equation}
({\cal P}_\sigma f)(\phi) = f_\sigma e^{i \sigma \phi}.
\end{equation}
Hence the projection operator for $\rho_j \otimes \rho_\sigma$ is 
\begin{equation}
{\cal P}_{j,\sigma} = {\cal P}_j \otimes {\cal P}_\sigma \,.
\end{equation}

To exhibit the twist, we must consider fields $\psi_0$ which are functions in ${\mathbb Z}_n \otimes {\mathbb Z}$. 
So a point $p$ is now a pair $(k,\sigma)$. We can now write the twist of the multiplication operator from $\Delta_0$ to $\Delta_\theta$, $\theta$ being a function on $({\mathbb Z}_m \times {\mathbb Z}) \times ({\mathbb Z}_{m^\prime} \times {\mathbb Z})$, with values $\theta_{i \sigma,j \tau}$. Then 
\begin{equation}
\Delta_\theta = {\cal P}_{i \sigma} e^{i/2 m_{i \sigma} \theta_{i \sigma,j \tau} m^\prime_{j \tau}} {\cal P}_{j \tau}
\end{equation}
with summation on repeated indices and $\theta_{i \sigma,j \tau} = -\theta_{j \tau,i \sigma}$.

As regards applications, we can think of the following. The group $G$ is the mapping class group of an asymptotically flat spatial slice, and affects $\psi_0$ only in the region where the geon is localized. The twist $F_\theta$ also has that property. Hence it affects only Planck-scale physics. With the modification of $\psi_0$ to $\psi_\theta$, we can then construct phenomenological models to probe spacetimes at Planck scale. An application along these lines to CMB radiation and FRW metric will be discussed below. For reviews of the standard FLRW model, see  \cite{Guth:1980zm,Linde:1981mu,Albrecht:1982wi}. For our work, see \cite{Akofor:2007fv,Akofor:2008gv,Joby:2014oee}.

The analogs of mapping class groups proliferate in the topology of extended objects like one or several closed strings. In that context, they are called motion groups \cite{goldsmith}. Long ago, we discussed the applications of motion groups and the emergence of exotic statistics therefrom in physics \cite{Aneziris:1990gm}.  

There are also applications to particle physics. The groups $G$ and $A$ in the above discussion could have been Lie groups. It is only that $A$ has to be abelian. That gives us the possibility of twisting the flavour symmetry of $QCD$ say and examining its phenomenology \cite{Balachandran:2006hr,Srivastava:2012hn}.

\section{CMB Anisotropies and Noncommutative Geometry}

In the current ''standard" model, CMB anisotropies are sourced by quantum fluctuations $\hat{\rho}$ of the field $\phi$ which drives the inflation.   The power spectrum $P_\rho( \vec{k}, \eta)$ at momentum $\vec{k}$ and conformal time $\eta$, defined by  
\begin{equation}
\langle 0 | \hat{\phi}(\vec{k},\eta) \hat{\phi}(-\vec{k}',\eta) |0 \rangle = (2 \pi)^3 P_\phi( \vec{k}, \eta) \delta^3(\vec{k}-\vec{k}')
\end{equation}
then causes fluctuations of the gravitational field which are seen by current observations. For a review of these results, adapted to this article, we refer to the articles \cite{Guth:1980zm, Linde:1981mu, Albrecht:1982wi}.
	
The background metric in these calculations is the FLRW metric in conformal time $\eta$:
\begin{equation}
ds^2 = a^2(\eta) (dt^2 - d\vec{x}^2),
\end{equation}
while $|0\rangle$ is the ground state of $\phi$.

For Minkowski metric which is invariant under all spacetime translations, the Fock space annihilation-creation operators $c_{\vec{p}}, c_{\vec{p}}^\dagger$ get twisted to $a_{\vec{p}}, a_{\vec{p}}^\dagger$ on the Groenewold-Moyal plane as we have seen, where
\begin{equation}
a_{\vec{p}}=c_{\vec{p}}e^{-i/2 p \wedge P}, \quad a_{\vec{p}}^\dagger  = c_{\vec{p}}^\dagger e^{+i/2 p \wedge P},
\quad p \wedge P = p_\mu \theta_{\mu \nu} P_\nu \,.
\label{tw1}
\end{equation}
An important consequence is that$[a_{\vec{p}},a_{\vec{q}}] \neq 0$ for $\vec{p} \neq \vec{q}$:
\begin{equation}
a_{\vec{p}} a_{\vec{q}} = c_{\vec{p}}e^{-i/2 p \wedge q}c_{\vec{q}}e^{-i/2 (p+q) \wedge P} \neq a_{\vec{q}} a_{\vec{p}} \,. 
\end{equation}
We will revisit this equation shortly as also the fact that $\phi_\theta$ is no longer Gaussian (quasi-free) even on Minkowski space. But for now, we focus on the effect of twist on the CMB spectrum.

We retain the twist in \eqref{tw1} also for the FLRW metric, and denote the twisted ground state and operators by the 
same symbols. Since the FLRW metric is invariant under spatial translations, so is the new ground state 
$|0\rangle$. Then by spatial momentum conservation, expectation values such as $\langle 0 | a_{\vec{p}} a_{\vec{q}} |0\rangle$ vanish unless $\vec{p} + \vec{q} = 0$. But if $\vec{p} + \vec{q} = 0$, $p \wedge q$ 
becomes $p_0 \theta_{0i} q_i + p_i \theta_{0i} q_0= 2p_0 \theta_{0i} q_i$. Calling $(\theta_{01}, \theta_{02},\theta_{03})$ as $\vec{\theta}$, this is $2p_0 \vec{\theta} \cdot \vec{q} P_0$ on $|0\rangle$.

The CMB fluctuations are supposed to be sourced by the power spectrum $P_\rho$. It gets coupled to metric perturbations and since they influence the photons, lead to the observed fluctuations. Thus the basic quantity of interest is the two-point function of the field $\phi_\theta$.

Let us briefly consider the $n$-point function of $\phi_\theta$ in the Minkowski vacuum (See \cite{Akofor:2007fv,Akofor:2008gv,Joby:2014oee} for our work). We will see that it is not Gaussian correlated and leads the 
way to the calculation of the modified power spectrum for the FLRW metric. 

Using $\phi_\theta = \phi_0 e^{\frac{i}{2} \overleftarrow{\partial} \wedge P}$, we get, for Minkowski vacuum,
\begin{equation}
\langle 0| \phi_\theta(x_1) \phi_\theta(x_2) \cdots \phi_\theta(x_n) |0 \rangle = \langle 0| \phi_0(x_1) \phi_0(x_2) \cdots \phi_0(x_n) e^{-\frac{i}{2} \sum_{J=1}^n \sum_{I=2}^{J-1} \overleftarrow{\partial}_{x_I} \wedge \overleftarrow{\partial}_{x_J}} |0 \rangle \,.
\end{equation}
 Setting
 \begin{equation}
 \phi_\theta(x) = \phi_\theta(\vec{x},t) = \int \frac{d^3k}{(2\pi)^3} \Phi_\theta (\vec{k},t) e^{i \vec{k} \cdot \vec{x}},
 \end{equation}
 we find 
 \begin{multline}
\langle 0| \Phi_\theta(\vec{k}_1,t_1) \Phi_\theta(\vec{k}_2,t_2) \cdots \Phi_\theta(\vec{k}_n,t_n) |0 \rangle = \\
 \exp \left(\frac{i}{2} \sum_{J>I}(\vec{k}_I \wedge \vec{k}_J) \right) 
\langle 0| \phi_0(\vec{k}_1, t_1 + (\vec{\theta}_0 \cdot \vec{k}_2 + \cdots \frac{\vec{\theta}_0 \cdot \vec{k}_n}{2})) \times \\
\phi_0(\vec{k}_2, t_2 -\vec{\theta}_0 \cdot \vec{k}_1 + \frac{\vec{\theta}_0 \cdot \vec{k}_3 +\cdots \vec{\theta}_0 \cdot \vec{k}_n}{2}) \times \\
\cdots \phi_0(k_n, t_n+ \frac{-\vec{\theta}_0 \cdot \vec{k}_1 - \vec{\theta}_0 \cdot \vec{k}_2 \cdots -\vec{\theta}_0 \cdot \vec{k}_{n-1} + \vec{\theta}_0 \cdot \vec{k}_n}{2}) |0 \rangle \,. 
\end{multline}

The FLRW has spatial translational symmetry so that we may set $\sum_i \vec{k}_i=0$. Hence the $n$-point function becomes
\begin{multline}
\langle 0| \Phi_\theta(\vec{k}_1,t_1) \Phi_\theta(\vec{k}_2,t_2) \cdots \Phi_\theta(\vec{k}_n,t_n) |0 \rangle \\
= \exp \left(\frac{i}{2} \sum_{J>I}(\vec{k}_I \wedge \vec{k}_J)\right) \langle 0 |\Phi_0 (\vec{k}_1,t_1 - \frac{\vec{\theta}_0 \cdot \vec{k}_1}{2}) \Phi_0 (\vec{k}_2 ,t_2 - \vec{\theta}_0 \cdot \vec{k}_1 - \frac{\vec{\theta}_0 \cdot \vec{k}_2}{2}) \\ \cdots \Phi_0 (\vec{k}_n, t_n - \vec{\theta}_0 \cdot \vec{k}_1 - \vec{\theta}_0 \cdot \vec{k}_2 \cdots - \vec{\theta}_0 \cdot \vec{k}_{n-1}) | 0 \rangle \,.
\label{npointfn}
\end{multline}

For the two-point function of our interest, then,
\begin{equation}
\langle 0| \Phi_\theta(\vec{k}_1,t_1) \Phi_\theta(\vec{k}_2,t_2) |0 \rangle = \langle 0| \Phi_0 (\vec{k}_1,t_1 - \frac{\vec{\theta}_0 \cdot \vec{k}_1}{2} \Phi_0 (- \vec{k}_1, t_2 - \frac{\vec{\theta}_0 \cdot \vec{k}_1}{2}) |0 \rangle
\label{twopointfn}
\end{equation}
since $\vec{k}_i \wedge \vec{k}_2 =0$ from $\vec{k}_1 + \vec{k}_2 =0$.

The expressions \eqref{tw1} and (\ref{twopointfn}) do not decompose into a sum of products of two-point functions and hence is are Gaussian correlated even for Minkowski spacetime as claimed.

The modified power spectrum has to be deduced from (\ref{twopointfn}) at equal times, $t_2=t_1$. But unlike the situation when $\vec{\theta} =0$, it is not real. We overcome this problem by replacing the product of fields at $t_2=t_1$ by their anti-commutator divided by two. Further, we replace the time arguments of fields by the conformal time $\eta$. The noncommutative power spectrum $P_{\Phi_{\theta}}(\vec{k}, \eta)$ is then given by
\begin{equation}
\frac{1}{2}\langle 0 | [\Phi_\theta (\vec{k}, \eta) \Phi_\theta (\vec{k}', \eta)]_+ |0 \rangle = (2 \pi)^3 P_{\Phi_0} (\vec{k}, \eta) \delta^3 (\vec{k} + \vec{k}')
\end{equation}

We refer to the articles \cite{Akofor:2007fv, Akofor:2008gv} and that of Joby et. al. \cite{Joby:2014oee} for the derivation of the CMB anisotropies from here and their comparison with data. The second paper has the most detailed comparison. If $H$ is the Hubble constant, the data constrain $H |\vec{\theta}||\vec{k}|$. That gives the lower bound of 20 TeV for the scale of the onset of spacetime noncommutativity: $|\vec{\theta}|^{1/2} \geq 20$ TeV.

\section{Moyal versus Wick-Voros}

The twist factor $F_\theta$ in the deformation leading to the Moyal algebra is only one choice for the twist. There are many other possible choices, the only constraint being that the deformed algebra be associative. Even this constraint perhaps can be discarded, requiring of the deformed product $*$ only that $A*(B*C)$ and $(A*B)*C$ are related say be a unitary transformation. But such choices are not much used in the literature. A generic method for finding new $*$'s are described in \cite{balbook}.

We want to discuss the Wick-Voros deformation which emerges from coherent states in this section \cite{Mario, review}. To distinguish it from the Moyal case, we will denote the product and the twist for the latter by $*_M$ and $F_\theta^M$ and for Wick-Voros by $*_V$ and $F_\theta^V$. We will argue that they are both $*$-isomorphic as Hopf algebras, and carry compatible twisted Poincar\'e group actions as well. All the same, this equivalence fails in quantum theory. The reason is that the Wick-Voros algebra is incompatible with the Hilbert space adjoint $\dagger$. This statement will be checked explicitly. 

Changing the previous notations, we will call the Fock space creation and annihilation operators for mass $m$ and momentum $p$ as $c_p^\dagger$ and $c_p$ and their twisted versions for Moyal and Wick-Voros will be $a_p^{M,V \dagger}$ and $a_p^{M,V}$.

We have seen that 
\begin{eqnarray}
a_p^{M *} &=& c_p^\dagger e^{-\frac{i}{2} p_\mu \theta_{\mu \nu} P_\nu} = e^{-\frac{i}{2} p_\mu \theta_{\mu \nu} P_\nu} c_p^\dagger, \\
a_p^{M} &=& c_p e^{\frac{i}{2} p_\mu \theta_{\mu \nu} P_\nu} = e^{\frac{i}{2} p_\mu \theta_{\mu \nu} P_\nu} c_p.
\end{eqnarray}
Thus $a_p^M$ is the Hilbert space adjoint of $a_p^{M*}$: $(a_p^{M*})^\dagger = (a_p^{M*})^* = a_p^M$.

The twisted versions $a_p^{V*}$ and $a_p^{V}$ are as follows:
\begin{eqnarray}
a_p^{V*} &=& (a_p^{M})^* e^{-\theta \,p_\mu  P^\mu} \label{v1}\\
a_p^V &=& (a_p^{M})e^{\theta \,p_\mu  P^\mu} \label{v2}
\end{eqnarray}
where we have specialized to two dimensions and $\theta_{\mu \nu} = \epsilon_{\mu \nu}$.

In the Moyal case, the twisted field $\phi_\theta^M$ can be obtained from $\phi_0^M$ by an overall twist: $\phi_\theta^M = \phi_0^M e^{-\frac{i}{2} \overleftarrow{P}_\mu \theta_{\mu \nu} \overrightarrow{P}_\nu}$ as we saw. An overall twist works also 
for Wick-Voros, $a_p^{V*}e^{-i p\cdot x} = a_p^{M*}e^{-i p \cdot x} e^{-i \theta \overleftarrow{\partial} \cdot \overrightarrow{P}}$ and $a_p^{V}e^{i p\cdot x} = a_p^{M}e^{i p \cdot x} e^{-i \theta \overleftarrow{\partial} \cdot \overrightarrow{P}}$. So $\phi_\theta^V = \phi_\theta^M e^{-i \theta \overleftarrow{\partial} \cdot \overrightarrow{P}}$.

But $a_p^{V*} \neq (a_p^V)^\dagger$ where $\dagger$ is the Hilbert space adjoint of $a_p^V$ as one sees from (\ref{v1},\ref{v2}). Thus there is a quantum anomaly for the Wick-VOros at the Hilbert space level. $a_p^{M,V}$ and their adjoints are related by an overall twist of fields. That would not be the case if we had set $(a_p^V)^* = (a_p^V)^\dagger$ since 
\begin{equation}
(a_p^{V*})^\dagger = e^{-\theta p \cdot P} a_p^M = a_p^M e^{\theta p \cdot p} e^{-\theta p \cdot P} \neq a_p^V
\end{equation}
Thus there is a quantum anomaly at the Hilbert space level for Wick-Voros.

But there is no such anomaly at the classical algebraic level. For the algebra of functions on Minkowski space, the Moyal and Wick-Voros products are
\begin{eqnarray}
f_1 *_M f_2 &=& m_0 [F_\theta^M(f_1 \otimes f_2)], \\
f_1*_V f_2 &=& m_0[F_\theta^V(f_1 \otimes f_2)]
\end{eqnarray}
where 
\begin{eqnarray}
F_\theta^M &=& e^{\frac{i}{2} \partial_\mu \theta_{\mu \nu} \otimes \partial_\nu} \\
F_\theta^V &=& F_\theta^M e^{\theta \partial \otimes \partial^\mu}
\end{eqnarray}
At the classical level, the Moyal and Voros-Wick algebras are also $*$-isomorphic as Hopf algebras. Thus there is an invertible map $T$ from ${\cal A}_\theta^M$ to ${\cal A}_\theta^V$ such that 
\begin{eqnarray}
T: f &\rightarrow&  (Tf), \\
T: f^* &\rightarrow& (Tf)^* 
\end{eqnarray}
and
\begin{equation}
T m_0 F_\theta^M (f_1 \otimes f_2) = m_0 F_\theta^V (Tf_1 \otimes Tf_2)
\end{equation}
The map $T$ is simple to write down:
\begin{equation}
T=e^{-\frac{\theta}{4} \nabla^2}
\end{equation}
We can verify that $T$ establishes the above $*$-isomorphism by using plane waves for $f$ and $f_i$:
\begin{eqnarray}
(Te_p)(x) &=& e^{-\frac{\theta}{4} \nabla^2} e^{i p \cdot x} = e^{\frac{\theta}{4}p^2} e_p(x), \\
T m_0 (F_\theta^M e_p \otimes e_q) &=& T e_{p+q} e^{i p \wedge q} = e^{\frac{\theta}{4}(p+q)^2} e_{p+q} e^{i \theta p \wedge q},
\end{eqnarray}
and
\begin{equation}
m_0 (F_\theta^V (Tf_1 \otimes Tf_2) = e^{i \theta p \wedge q} \, e^{\frac{\theta}{2} p \cdot q}e^{\frac{\theta}{4}(p^2+q^2)}e_{p+q} = e^{\frac{\theta}{4}(p+q)^2} e_{p+q} e^{i \theta p \wedge q}
\end{equation}
showing the claimed classical isomorphism.

\begin{figure}[!htb]	
\centering
	\includegraphics[width=0.5\textwidth, height=0.20\textheight]{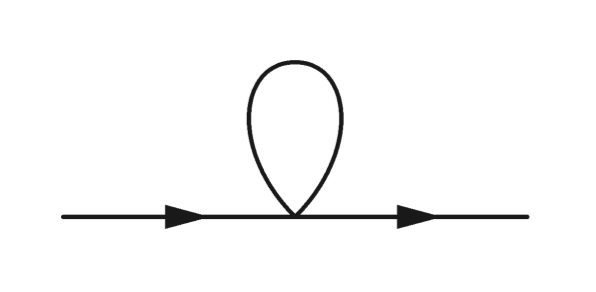}
	\caption{Self Enegy Diagram}
	\label{fig1}
\end{figure}

The quantum anomaly shows up in calculations. Thus consider the self-energy diagram in figure \eqref{fig1} for the $(\phi*_{M,V})^4$ interaction for a massive scalar field. 
The loop gives $\int \frac{d^4k}{k^2+m^2}$ for Moyal, but $e^\frac{p^2}{4} \int \frac{d^4 k}{k^2+m^2} e^{\frac{\theta}{4}k^2}$ for Wick-Voros. 

Enough has been said in this short review to establish the inequivalence of the two deformations. The original papers contain further details (See \cite{Mario2} and especially \cite{Mario3}).

\section{The Euclidean Formulation for Moyal $*$}

Standard fields like $\phi_0$, with finite mass $m$ have an analytic continuation in time $t$ to imaginary values, $t=-i \tau$, with $\tau >0$. They were first studied by Schwinger. Later their properties were formalized by Osterwalder and Schrader (OS) \cite{Osterwalder:1973dx,Osterwalder:1974tc}. If their axioms are fulfilled, then the Minkowski space theory can be recovered from the Euclidean space correlation functions.

A particularly important axiom of OS is reflection positivity. It is implied by the positivity of the real time scalar product, a property often called unitarity. We will explain this axiom below.

Euclidean correlation functions are often described using functional integrals. In this short discussion below, we show that the treatment of reflection positivity involves special features on the Moyal plane. We do not treat the related functional integral here.

There is no loss of content in restricting to $(1+1)$ spacetime. So $\phi_\theta$ here is $\phi_\theta(x,t)$. We will smear $\phi_\theta$ with spatial real test functions which vanish fast at spatial infinity: $f \in {\cal D}^\infty$. Thus we consider 
\begin{equation}
\phi_(f,t) = \int \phi_\theta(x,t) f(x)
\end{equation}
Consider first $\phi_0$, a real scalar field of mass $m$. We have that $\phi_0(f,t) = e^{i H t} \phi_0 e^{-i H t}$.  Continuing $t$ to $-i \tau$ we get 
\begin{equation}
\phi_0(f, -i \tau) = e^{H \tau} \phi_0(f) e^{-H \tau}
\end{equation}
Hence,
\begin{equation}
\phi_0(f,-i \tau)^* = e^{-H \tau} \phi_0)(f) e^{H \tau} = \phi_0 (f,+i \tau)
\end{equation}
Therefore, hermitian conjugation is equivalent to the "reflection" $\tau \rightarrow -\tau$.

Now let $\phi_0(f_i,\tau_i), i = 1,\cdots N$ be $N$ such Euclidean $\phi_0$. Consider 
\begin{equation}
\phi_0(f_1, -i \tau_1) \phi_0(f_2,-i \tau_2) \cdots \phi_0(f_n,-i\tau_N) |0\rangle \,,
\end{equation}
where $|0\rangle$ is the vacuum vector. 

The scalar product of this with its adjoint must be positive:
\begin{multline}
\langle0| \phi_0(f_N,-i\tau_N)^* \phi_0(f_{N-1},-i\tau_{N-1})^* \cdots \phi_0(f_1,-i\tau_1)^* \\
\phi_0(f_1,-i\tau_1) \phi_0(f_2,-i\tau_2) \cdots \phi_0(f_N,-i\tau_N) |0\rangle >0
\end{multline}
Hence
\begin{multline}
\langle0|\phi_0(f_N,+i\tau_N) \phi_0(f_{N-1},+i\tau_{N-1}) \cdots \phi_0(f_1,i\tau_1) \\
\phi_0(f,-i\tau_1) \phi_0(f_2, -i\tau_2) \cdots \phi_0(f_N,-i\tau_N) |0\rangle > 0
\end{multline}
This is reflection positivity, also for interacting fields. 

Now consider the twisted field $\phi_\theta$. We have
\begin{equation}
\phi_\theta(x,t) = \phi_0(x,t) e^{\frac{i}{2} \theta (\overleftarrow{\partial_x} \overrightarrow{\partial_t} -  \overleftarrow{\partial_t} \overrightarrow{\partial_x})}
\end{equation}
Continuation to Euclidean time gives 
\begin{equation}
\phi_\theta(x,-i \tau) = \phi_0(x,-i\tau) e^{-\frac{1}{2} \theta ( \overleftarrow{\partial_x} \overrightarrow{\partial_\tau} -  \overleftarrow{\partial_\tau} \overrightarrow{\partial_x})}
\end{equation}
so that
\begin{equation}
\phi_\theta(x,-i\tau)^* = \phi_0(x, i \tau) e^{\frac{1}{2} \theta (\overleftarrow{\partial_x} \overrightarrow{\partial_\tau} -  \overleftarrow{\partial_\tau} \overrightarrow{\partial_x})} = \phi_\theta(x,+i \tau).
\end{equation}
Thus, we still have reflection positivity. The exponent here has changes sign because for example
$\phi_\theta(x,-i\tau) e^{\frac{\theta}{2}(\overleftarrow{\partial_x} \overrightarrow{\partial_\tau} -  \overleftarrow{\partial_\tau} \overrightarrow{\partial_x})} \phi_\theta(x,-i\tau)$ under $*$ is $\phi_\theta (x,i\tau) e^{-\frac{\theta}{2}(\overleftarrow{\partial_x} \overrightarrow{\partial_\tau} -  \overleftarrow{\partial_\tau} \overrightarrow{\partial_x})} \phi_\theta(x,i\tau)$.

But if $\tilde{\phi}_\theta(x,\tau) = \phi_\theta(x,-i\tau)$ and we write $\tilde{\phi}_0(x,\tau) * \tilde{\phi}_0(x,\tau) = \phi_0(x,\tau) e^{\frac{i}{2} \theta (\overleftarrow{\partial_\tau} \overrightarrow{\partial_x} -  \overleftarrow{\partial_x} \overrightarrow{\partial_\tau})} \tilde{\phi}_0(x,\tau)$,
then we pick up an extra $i$ in the exponential, resulting in the loss of reflection positivity.

\section{Discussion and Outlook}

Fundamental spacetime noncommutativity is expected to appear typically at Planck scale, but emergent or effective noncommuting coordinates can also make their appearance in systems like quantum Hall effect. We have provided a description of our approach to field theories on such spaces, which keeps central the idea of causality, and to a large extent, also a generalized notion of locality. Maintaining the principle of causality and (twisted) locality in a manner consistent with the underlying noncommutative spacetime algebra leads to a deformation of the connection between spin and statistics. Observation of such a deviation from the standard spin-statistics relation may be one of the clearest signals of quantum gravity. Whether such delicate and subtle experiments can be conducted remains to be seen, and is perhaps the foremost challenge facing theoretical models of fields on noncommutative spacetimes.

\end{document}